\documentclass{article}
\usepackage{authblk}
\usepackage[symbol]{footmisc}

\title{Synergistic effect of the electronic band delocalization and bond anharmonicity on the thermoelectric performance of Cs$_2$TeX$_6$ (X=Cl, Br, I)}
\author[1]{Heena$^{\dagger}$}
\author[2]{Vineet Kumar Pandey$^{\dagger}$}
\author[3]{Saanvi Marethiya}
\author[4]{Ambesh Dixit}
\author[5]{Ajay Singh Verma$^*$}
\author[3]{K.C. Bhamu$^*$}%\thanks{Corresponding author}\thanks{Email: kcbhamu85@gmail.com}}
\footnotetext[2]{These authors contributed equally to this work}
\footnotetext[1]{Corresponding author. Email: kcbhamu85@gmail.com}
\affil[1]{Department of Physics, Jamia Millia Islamia, New Delhi 110025, India}
\affil[2]{Department of Physics, Govt Naveen College Janakpur, M.C.B. 497778, Chhattisgarh, India}
\affil[3]{Department of Physics, SLAS, Mody University of Science and Technology, Lakshmangarh, Sikar, Rajasthan, 332311, India}
\affil[4]{Advanced Materials and Device Laboratory, Department of Physics, Indian Institute of Technology, Jodhpur 342037, India}
\affil[5]{Department of Physics, Anand School of Engineering and Technology, Sharda University Agra, Keetham, Agra, 282007}
\date{\today}
\usepackage{gensymb}
\usepackage{amsmath}
\usepackage{upgreek}
\usepackage{amssymb}
\usepackage{amsfonts}
\usepackage{graphicx,multirow,color}
\usepackage{subcaption}
\usepackage[defaultcolor=red]{changes}
\usepackage{float}
\usepackage{titlesec}
\usepackage{pdfpages}
\usepackage{setspace}
\usepackage{cite}
\DeclareUnicodeCharacter{200B}{}
\titleformat{\paragraph}
{\normalfont\normalsize\bfseries}{\theparagraph}{1em}{}
\titlespacing*{\paragraph}
{0pt}{3.25ex plus 1ex minus .2ex}{1.5ex plus .2ex}
\usepackage{geometry}
\begin{document}
\maketitle

%\tableofcontents
\newpage
\begin{figure}[ht]
 \centering
 \includegraphics[width = 0.9\textwidth]{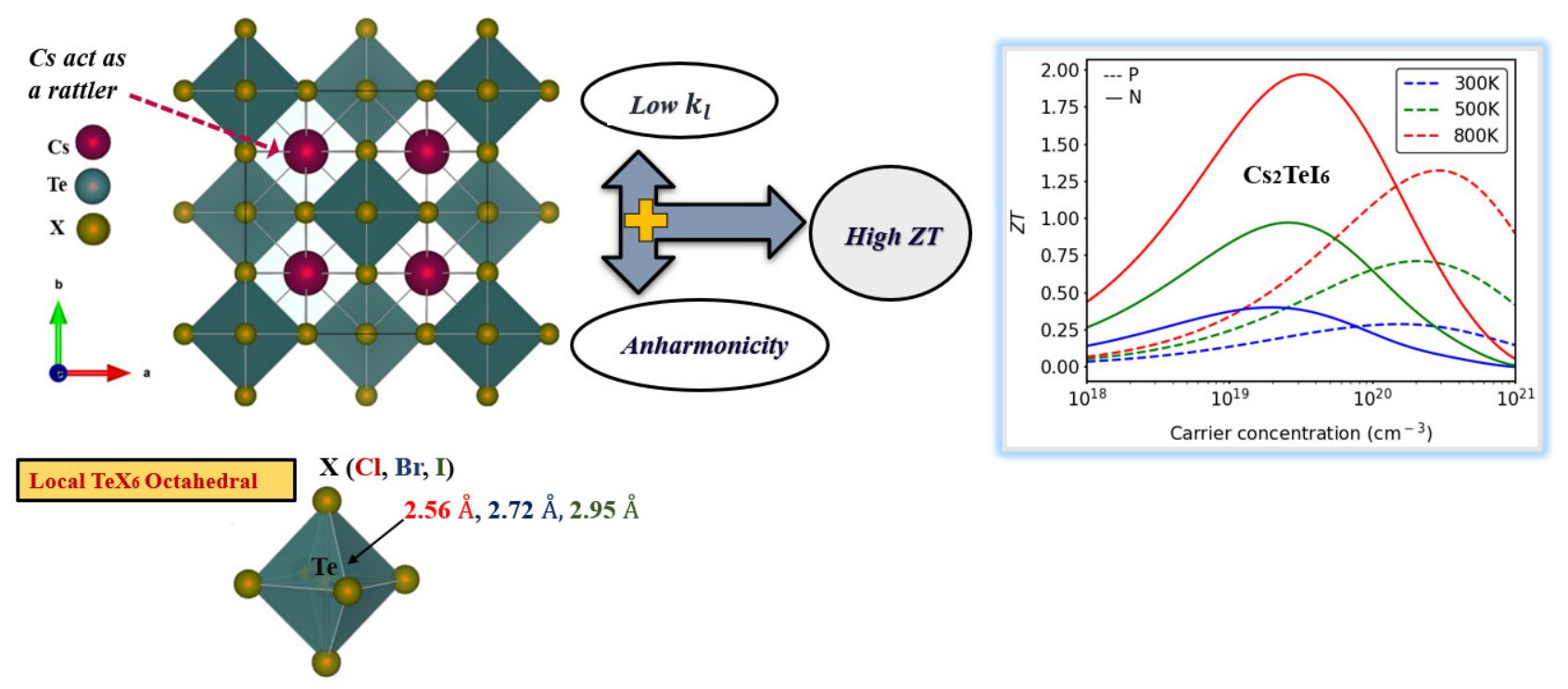}
%\caption{}
                  \label{toc}
\end{figure} 
\newpage
\paragraph*{keywords}
\begin{itemize}
\item Vacancy ordered double halide perovskites
\item Ultralow lattice thermal conductivity
\item Inverse participation ratio
\item High power factor and ZT
\end{itemize}
\newpage
\section{Abstract}
We investigate the structural, mechanical, and the thermoelectric properties of lead-free double halide perovskites Cs$_2$TeX$_6$ (X=Cl, Br, I) by employing first-principles calculations and semiclassical Boltzmann transport theory. The HSE06 band gap is incorporated using scissor correction method along with PBE calculated electronic band structure including spin-orbit coupling (SOC) to accurately predict the transport properties. The band gap values are 3.27, 2.50 and 1.55 eV for Cs$_2$TeX$_6$ (X=Cl,Br, I) respectively. The coexistence of heavy and light bands in the case of Cs$_2$TeI$_6$ band structure helps to mitigate  trade-off between the Seebeck coefficient (S) and electrical conductivity ($\sigma$). Among these systems, Cs$_2$TeI$_6$ exhibits superior performance with a remarkable ZT of 1.97 at 800 K and electronic concentration $3.35 \times 10^{19}$ cm$^{-3}$. Such a high ZT at significantly low carrier concentration is the result of high electrical conductivity along with low lattice thermal conductivity. The lattice thermal conductivity ($\kappa_l$) of Cs$_2$TeI$_6$ was found to be 0.41 Wm$^{-1}$K$^{-1}$ at room temperature. Such a low $\kappa_l$ is attributed to the presence of weak Te-I bond and non-uniform out-of-phase displacement of Cs atoms. The presence of local octahedra TeX$_6$ along with weak bonds offers strong resistance to heat conduction and, in turn, lattice thermal conductivity is suppressed drastically. In particular, transverse acoustic phonons along with optical phonons significantly limit the lattice thermal conductivity. Taken together, these results position Cs$_2$TeI$_6$ as a strong candidate for high performance thermoelectric applications. 
 
\section{Introduction}
The rapid requirement in global energy consumption is driving fossil fuel exhaustion at a very high pace and causing severe damage to the environment and global warming. In response, scientists are more focused on adopting renewable energy resources and efficient use of energy to avoid energy losses. One promising way out is the reuse of waste heat generated by primary energy resources.
Thermoelectrics have emerged as a clean source of electrical energy by utilizing heat as an input \cite{thm1, thm2, thm3, thm4}. Thermoelectric materials promise many advantages such as no waste output, no moving parts involved and it alleviates further pollution increase by reusing the waste energy \cite{thm2}. Conventionally, a thermoelectric material's conversion efficiency is defined by figure of merit (ZT)= $\frac{S^2 \sigma T}{\kappa_e+\kappa_l}$, in which S, $\kappa_e$, $\kappa_l$, $\sigma$ and T are the Seebeck coefficient, electronic thermal conductivity, lattice thermal conductivity, electrical conductivity and absolute temperature, respectively \cite{zt1, zt2}. In principle, high   Seebeck coefficient, low thermal conductivity and high electrical conductivity ($\kappa = \kappa_e  + \kappa_l$) need to be possessed by a material in order to be highly efficient in energy conversion.  Since there is a tradeoff between S and $\sigma$ on the scattering mechanism and carrier concentration, they cannot be increased simultaneously, i.e., the increase in S is followed by decrease in $\sigma$ and vice versa. Apart from this, another approach to make these materials energy efficient is by hindering the phonon transport using defects, scattering or doping \cite{klstrategy, min1, min2, min3}. The latter approach of reducing phonon transport is easier as compared to enhancing electrical conductivity and the Seebeck coefficient. There is another route to look for a new class of materials that inherently accommodates high atomic weight, disorder introduced using dopants, the rattling motion or the polyhedra within the unit cell that enhance the phonon scattering and have intrinsic low lattice thermal conductivity thereby improving the efficiency of thermoelectric materials.    

The perovskite family offers many opportunities to overcome challenges posed to thermoelectric performance  by making use of its immense structural and diverse compositions \cite{pero1, inor1, pero2, pero3, pero4, pero5, pero6}.  These perovskites adopt the cubic structure with the chemical formula ABX$_3$, where A and B are positively charged cations, atom X is a halogen atom, and atom A is larger than B in size \cite{abx3}. Here atoms A and B are surrounded by anions in cuboctahedral and octahedral coordination, respectively. Later it was found that the size of A, B and X atoms significantly affects the stability and has high defect density \cite{stab1, stab2, stab3}. There are not many options available to tune the band gap of these perovskites \cite{bg1}. To overcome these, scientists introduced two cations to enable better size matching. Such structures have better stability, flexibility and better control over band gap tunability and are called double perovskite with chemical formula A$_2$BB$'$X$_6$. Double perovskites are a very versatile class of material including organic, inorganic or hybrid \cite{inor1, dhp1, dhp2, dhp3, dhp4, dhp5, dhp6}. Scientists have extensively exploited these materials for their optical, photochemical, transport, and various other properties \cite{dhp2, dhp3, dhp4, dhp5, dhp7, dhp8, dhp9, trdhp1, pdp}. In addition to being environmentally friendly and innocuous, these materials offer strong mechanical and dynamical stability, are easily synthesized, etc \cite{syn1, syn2, syn3}. These materials have been harnessed in light-emitting diode, solar cells and laser applications for their interesting properties \cite{laser1, laser2, laser3, solar1, solar2}.  In particular, lead halide perovskites have been reported to have low lattice thermal conductivity due to their unique lattice dynamics such as CsPbI$_3$, CsPbBr$_3$, CsSnI$_3$ \cite{cspbx3}. Although these systems provided very promising solution for thermoelectric applications, it was discarded for lead being toxic and Sn being unstable in the air at ambient conditions. Further, these perovskites do not exhibit better electrical transport properties as these have high band gaps (typically $\geq$ 1.8 eV) that restrict the electron conduction and pose a significant challenge \cite{bg_cspbx3}. Hence it is crucial to discover new material with better electrical properties and lower thermal conductivity. Lately, a distortion was introduced in the form of vacancy in double halide perovskite at the B site to make vacancy ordered double perovskite with chemical formula A$_2$BX$_6$ \cite{a2bx6, a2bx61, a2bx62}. The creation of the vacancy further improved the stability. Studies have suggested that accommodation of distortion in the local structure acts as a scattering point and significantly lowers the lattice thermal conductivity \cite{kla2bx6, klcs2sni6, klcs2bi6}. The obvious distinction between the structure of perovskites A$_2$BB$'$X$_6$ and A$_2$BX$_6$ is the local surrounding of the octahedra in A$_2$BX$_6$. The BX$_6$ octahedra share the corner in the former case, while it remains isolated in the latter case.  

Recently, a vacancy ordered double perovskite Cs$_2$TeI$_6$ has garnered significant attention due to its photovoltaic, optoelectronic properties, excellent absorption coefficient, stable X-ray detectors, energy storage applications and photodetector \cite{opto, xray, battery, photo}. Cs$_2$TeI$_6$ has been experimentally synthesized \cite{expt2, opto, expt}. Interestingly, the defect tolerance limit of this system Cs$_2$TeI$_6$ is significantly lower \cite{defect}. The band gap of this system is in the visible range (1.5-1.8 eV) \cite{defect}. %Though this system contains tellurium, it is mildly toxic and not yet identified as carcinogenic \cite{te}. 
In addition to this, Siad et al., showed the thermoelectric performance of Cs$_2$TeBr$_6$ and Cs$_2$TeI$_6$ at GGA-PBE level \cite{str}. This study did not thoroughly analysed the transport properties. Lately, Zheng et al predicted that Cs$_2$TeI$_6$ has ultralow lattice thermal conductivity \cite{kll}. In this work, we present the systematic study of structural, electronic, mechanical, vibrational and thermoelectric properties of Cs$_2$TeX$_6$ (X=Cl, Br, I).

\section{Computational methodology}
\label{cmp} 
Our first-principles calculations were based on the density functional theory as implemented in the Vienna Ab Initio Simulation Package (VASP) with the projector augmented wave (PAW) method \cite{vasp, vasp2, paww}. The exchange correlation potential of electrons was described by the Perdew-Burke-Ernzerhof (PBE) generalized gradient approximation (GGA) while optimizing the structure \cite{pbe}. As PBE underestimates band gap and to account for the fundamental self interaction error in PBE functional, we used Heyd-Scuseria-Ernzerhof (HSE06) method with an automatic $\Gamma$-centered k-mesh with resolution 0.025 $\textrm \AA^{-1}$ and 63 irreducible k-points for better prediction of band gaps and electronic properties \cite{hse}. We used an energy cutoff of 500 eV to truncate the plane wave basis used in Kohn-Sham wave functions. %The valence electron configurations of Cs, Te, Cl, Br, and I are  ------------ respectively. 
We optimized the structure until the Hellmann-Feynman force on each atom was less than 0.01 eV$\textrm \AA^{-1}$. We investigated the bonding analysis using the LOBSTER package \cite{lobster}. We computed the phonon dispersion using PHONOPY to confirm the dynamical stability of these systems \cite{phonopy}. We used a q-mesh $5 \times 5\times 5$ for phonon calculations. 
We computed the transport properties by solving the Boltzmann transport equation as implemented in AMSET code \cite{amset}. We plotted the electronic band structure and transport properties using SUMO code \cite{sumo}.  The structure and displacement modes were visualized using VESTA code \cite{vesta}. In order to calculate the lattice thermal conductivity, we used the Debye-Callaway model as incorporated in the AICON code \cite{aicon}. The input parameters needed to do this calculation such as phonon frequencies, group velocity and Gr\"uneisen parameter were computed using PHONOPY \cite{phonopy}. We calculated the Gr\"uneisen parameter using phonon calculation of optimized structure and 0.4$\%$ up and down in volume.

\section{Results and Discussion}
\subsection{Crystal structure optimization}
\label{strr}
\begin{figure}[H]
 \centering
 \includegraphics[width = 0.9\textwidth]{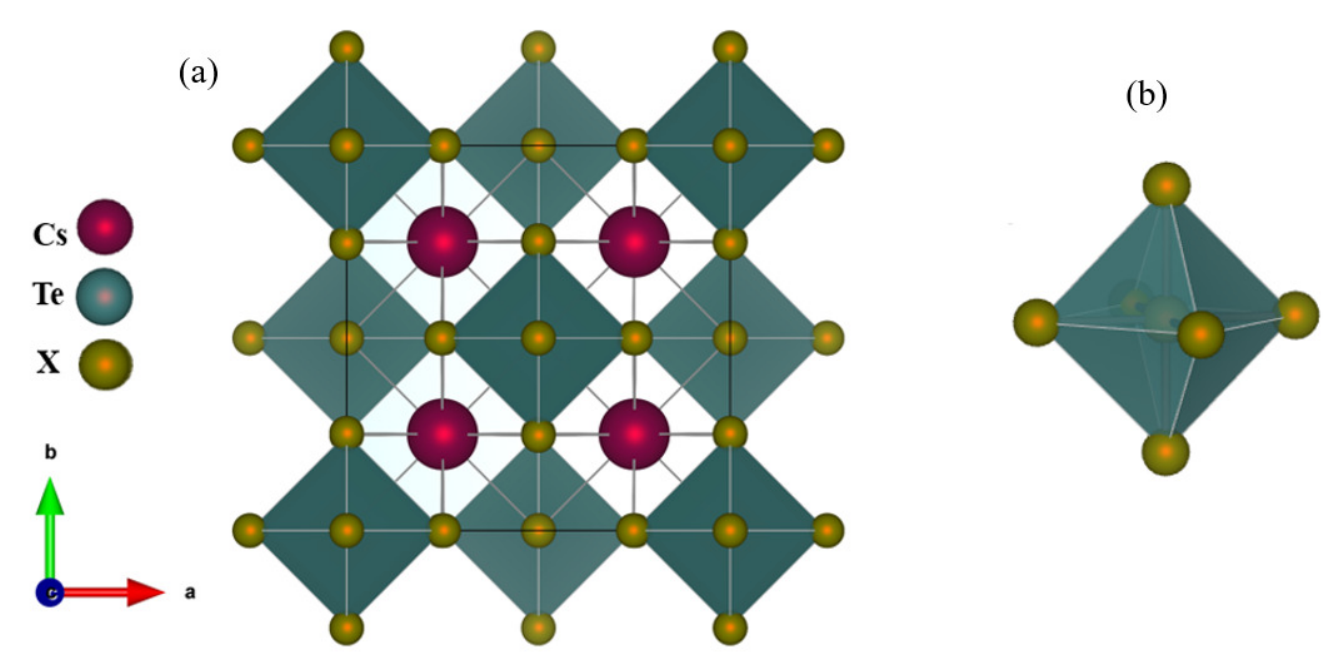}
 \caption{Crystal structure of vacancy ordered double halide perovskite (a) Cs$_2$TeX$_6$ (X = Cl, Br, I), (b) Octahedra structure (TeX$_6$)}
\label{cry}
\end{figure}

Cs$_2$TeX$_6$ (X=Cl, Br, I) are vacancy-ordered double halide perovskites that crystallize in a face-centered cubic (FCC) crystal structure with space group 
$Fm\overline{3}m$ \cite{str, experiment}. The crystallographic representation of the unit cell is illustrated in Fig. \ref{cry}(a) and (b). In this configuration, a unit cell of Cs$_2$TeX$_6$ accommodates [TeX$_6$]$^{-2}$ octahedra along with a cation Cs$^{+1}$ present in the void between the octahedra. These [TeX$_6$]$^{-2}$ octahedra do not share edges, faces or corners. The TeX$_6$ octahedra establish 12-fold coordination environment of X anions. Cs, Te, and X atoms are situated at 8c Wyckoff site with (0.25, 0.25, 0.25), 4a Wyckoff site with (0, 0, 0), and X anions at 24e Wyckoff site (0.2498, 0, 0) fractional coordinates, respectively. The Cs atom is embedded within the framework of TeX$_6$ octahedra. We have simulated the crystal geometry using GGA-PBE. Our calculations reveal that the Te-Cl, Te-Br and Te-I bonds are around 2.56 \AA, 2.73 \AA, and 2.95 \AA, \space respectively. The increasing bond length is attributed to decreasing electronegativity. The calculated lattice parameter of these systems Cs$_2$TeX$_6$ (X=Cl, Br, I) were found to be 10.51 \AA, 10.97 \AA, and 11.69 \AA, \space respectively. The increasing trend of lattice parameters is attributed to the increasing atomic radius of halides. These obtained values are in good agreement with the experimental and computational reports \cite{experiment, defect}. Due to the increasing volume of unit cell with increasing halides atomic number, the Cs-X bond length is found to increase to be 3.72 \AA, 3.88 \AA, and 4.13 \AA, \space for Cl, Br and I, respectively. In other words, the void space between the TeX$_6$ octahedra is increasing. The other structural details for Cs$_2$TeX$_6$ (X=Cl, Br, I) are provided in the table S1 of supplementary information.

\subsection{Bonding analysis} 
\label{cohps}
\begin{figure}[H]
 \centering
 \includegraphics[width = 0.9\textwidth]{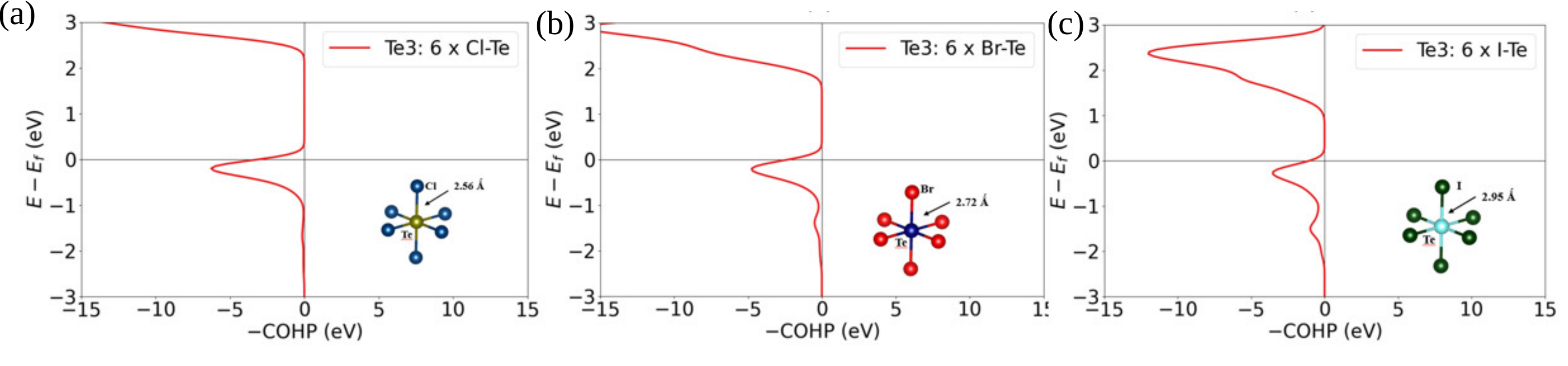}
 \caption{Projected crystal orbital Hamilton population of vacancy ordered perovskite derivatives for Cl-Te bond in (a) Cs$_2$TeCl$_6$ (b) Cs$_2$TeBr$_6$ (c) Cs$_2$TeI$_6$ }
\label{cohp}
\end{figure}

Figures \ref{cohp}(a), (b) and (c) display the projected crystal orbital Hamilton population (pCOHP) curves for the respective X-Te bonds in Cs$_2$TeX$_6$ (X=Cl, Br, I). The -pCOHP peak for other bonds is included in supporting information. The negative values on the X-axis represent antibonding states, with those below the Fermi energy indicating instability in the system. Our analysis of the -pCOHP curve reveals that with decrease in the electronegativity of the X atom, the heights of the -pCOHP peak decreases. However, we do not notice any significant shift of these peaks on the energy axis. This shortening of peaks is attributed to the diffused Iodine $p$ orbitals as compared to the Chlorine and Bromine $p$ orbitals. The decreasing heights of the peak imply reduction in orbital overlap, for instance, the I-Te has the least orbital overlap and the Cl-Te has the most orbital overlap. Orbital overlapping is related to the strength of bonds. In our case, I-Te bond is weaker than the Cl-Te and the Br-Te bond. Further, weak bonds generally correspond to more anharmonicity. Increased anharmonicity negatively impacts the lattice thermal conductivity and reduces it further \cite{anharmonic}. These primary results motivate to investigate further its thermoelectric properties.

\subsection{Electronic properties}
\begin{figure}[!h]
 \centering
 \includegraphics[width = 0.6\textwidth]{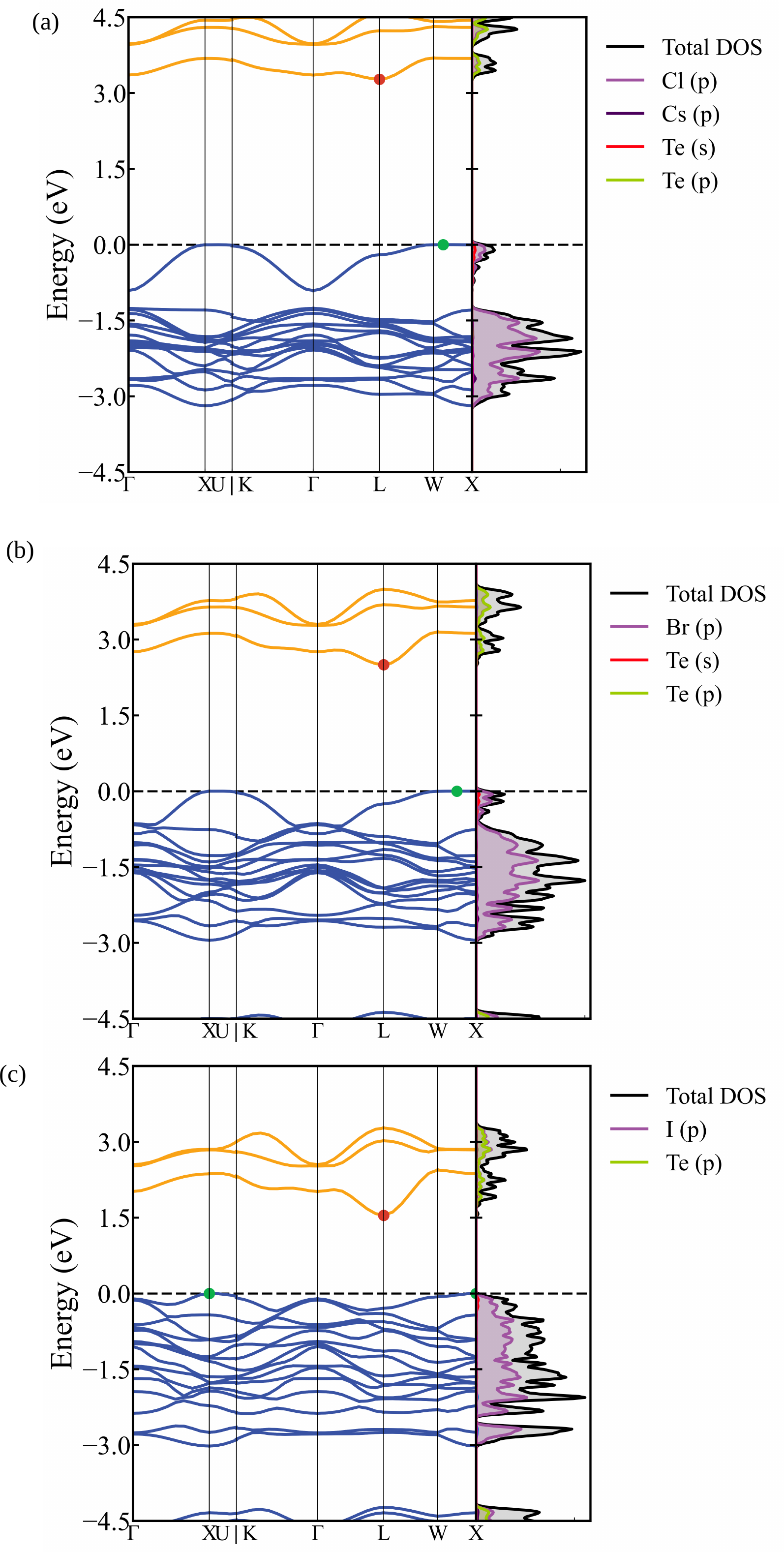}
 \caption{Electronic band structure along with DOS of (a) Cs$_2$TeCl$_6$ , (b) Cs$_2$TeBr$_6$, and (c) Cs$_2$TeI$_6$. }
\label{ebs}
\end{figure}

Figure \ref{ebs} depicts the electronic band structure along with the partial density of states (DOS) of Cs$_2$TeX$_6$ (X=Cl, Br, and I) using HSE06 hybrid functional including spin-orbit coupling. These materials are found to be semiconductors with electronic band gap values of 3.27 eV, 2.50 eV, and 1.55 eV, respectively at the HSE06 level. The decrease in band gap values is attributed to the decreasing electronegativities of Cl, Br, and I, and in turn increasing bond length of Te-Cl, Te-Br, and Te-I, respectively. Table \ref{tabb1} contains all calculated band gaps with different functionals along with respective experimental values from the reported literature. This matches well with the reported bandgaps in the available reports \cite{cmp1, bgcl, defect, opto}. The valence band maxima (VBM) and conduction band minima (CBM) of Cs$_2$TeCl$_6$ and Cs$_2$TeBr$_6$ lie at W and L high symmetry points of the Brillouin zone, respectively. However, the VBM of Cs$_2$TeI$_6$ shifts to $\Gamma$ point and CBM remains intact at L point. We do not notice any prominent change in the conduction band dispersion in Cs$_2$TeX$_6$ (X=Cl, Br, I), whereas there is significant improvement in the valence band dispersion. This modification in the band structure is present in DOS as well. Apart from this, the VBM is doubly degenerate in all the system Cs$_2$TeX$_6$ (X=Cl, Br, I). However, CBM is doubly degenerate in Cs$_2$TeCl$_6$ and Cs$_2$TeBr$_6$ and quadruply degenerate in Cs$_2$TeI$_6$. This higher degeneracy in CBM gives Cs$_2$TeI$_6$ an edge over the other two systems Cs$_2$TeCl$_6$ and Cs$_2$TeBr$_6$ from thermoelectric perspective. Table \ref{stdd} shows the effective mass of these systems along different directions of the Brillouin zone. Notably, there is presence of heavy holes and light holes together at the VBM for Cs$_2$TeCl$_6$ and Cs$_2$TeBr$_6$, while such feature in VBM of Cs$_2$TeI$_6$ is absent. The partial DOS analysis reveals that Te-p and X-p states have predominantly contributed to the valence band along with the minor contribution of Te-s states near VBM, while in the conduction band, Te-p states have dominant contribution. The significant dispersion of the conduction and valence band is attributed to the delocalized nature of $p$ orbitals. These combinations of different types of carriers have previously been reported to enhance the thermoelectric performance \cite{heavy, strategy}.  

\begin{table}[h]
\centering
 \caption{Calculated electronic Band gaps of Cs$_2$TeX$_6$ (X = Cl, Br, I) using different functionals}
\begin{tabular}{ c c c c c c} 
 \hline
VODPs & \multicolumn{3}{c}{Calculated band gap (eV)} & \multicolumn{2}{c}{Reported band gap (eV)} \\
 \hline 
        & PBE & HSE06 & HSE06+SOC & Theory & Experimental \\
          
 Cs$_2$TeCl$_6$ & 2.66 & 3.53 & 3.27    & 3.4   & 3.15 \cite{bgcl} \\
 
 Cs$_2$TeBr$_6$ & 1.97 & 2.72 & 2.50    & 2.7   & 2.68 \cite{bgcl}\\
 
 Cs$_2$TeI$_6$ & 1.18 & 1.85 & 1.55    & 1.83  & 1.59 \cite{defect, opto}\\
  \hline
\end{tabular}
\label{tabb1} 
\end{table}

\begin{table}[!hb]
\centering
 \caption{Carrier effective masses along different directions of the Brillouin zone}
\begin{tabular}{ c c c c c c c c c c c } 
 \hline

 System & \multicolumn{5}{c}{$m_h^*/m_e$} & \multicolumn{5}{c}{$m_e^*/m_e$} \\
 \hline
       & W-X ($\Gamma$ - X) & W-K ($\Gamma$ - K) & W-U ($\Gamma$ - L) & W-L & & & L-$\Gamma$ & L-U & L-W & L-K \\
  \hline
  Cs$_2$TeCl$_6$ & 10.442 & 2.986 & 2.986 & 1.577 & & & 1.446 & 0.643 & 0.643 & 0.643 \\
  
  Cs$_2$TeBr$_6$ & 15.491 & 2.624 & 2.624 & 1.302 & & & 0.702 & 0.380 & 0.380 & 0.380  \\
  
  Cs$_2$TeI$_6$  & (0.607, 0.839) & (0.614, 0.808) & (0.616, 0.798) & - & & & 0.388 & 0.224 & 0.224 & 0.224     \\ 
 \hline
\end{tabular}
\label{stdd} 
\end{table}

\subsection{Vibrational properties and electron localization function (ELF)}
\begin{figure}[H]
 \centering
 \includegraphics[width = 0.95\textwidth]{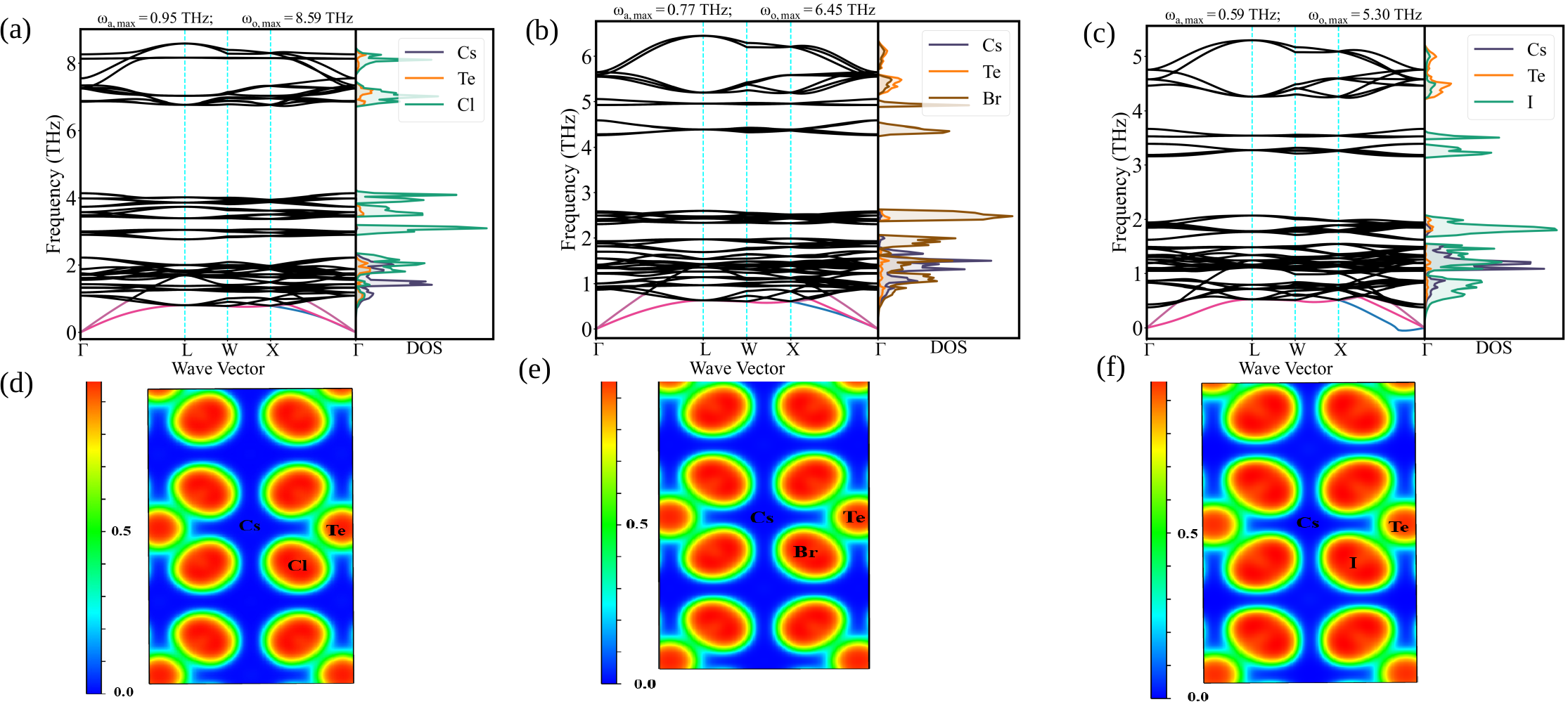}
 \caption{Phonon band structure along with DOS of (a) Cs$_2$TeCl$_6$ , (b) Cs$_2$TeBr$_6$, and (c) Cs$_2$TeI$_6$. Electron localization function of (110) plane passing through constituent atoms in (d) Cs$_2$TeCl$_6$ , (e) Cs$_2$TeBr$_6$, and (f) Cs$_2$TeI$_6$. The attached ELF scale of 0, 0.5, and 1.0 indicate electron depletion, delocalized and localized electron-like features, respectively.}
\label{phbs}
\end{figure}

Figures \ref{phbs} (a), (b), and (c) present the phonon band dispersion along the high-symmetry points of the Brillouin zones, along with the phonon partial density of states (PDOS) for Cs$_2$TeCl$_6$, Cs$_2$TeBr$_6$, and Cs$_2$TeI$_6$ respectively. We notice the presence of small imaginary frequency near the $
Gamma$ point in Cs$_2$TeI$_6$. However, Cs$_2$TeCl$_6$ and Cs$_2$TeBr$_6$ had no imaginary frequency. Such small imaginary frequencies are insignificant and can be overcome with larger supercells\cite{i1}. Hence, all these systems are dynamically stable. We notice the presence of a small imaginary frequency near $\Gamma$ point for Cs$_2$TeI$_6$. Such small imaginary frequencies are the results of computational inaccuracy. Going down in the halogen group from Cl to I, the unit cell accommodates more atomic mass and in turn, the vibrational frequencies soften. This softening leads to lower phonon energy and diminished Debye temperature thereby helping in reducing lattice thermal conductivity. For instance, the Debye temperature for Cs$_2$TeCl$_6$, Cs$_2$TeBr$_6$, and Cs$_2$TeI$_6$, estimated from their phonon band dispersion, are 412 K, 309 K, and 254 K, respectively. The atom projected phonon DOS reveals that the Cs atoms have a dominant contribution near the low frequency range. In the mid frequency range, the halogen atom has significant contribution with minor contribution from Te atoms. However, Te has mostly contributed near the high frequency range. Our comparison of phonon DOS reveals that the phonon DOS increases as we go from Cl to I, particularly at low range frequencies. Increasing the phonon DOS has previously been reported to cause higher phonon scattering and lower lattice thermal conductivity as low energy phonons are primarily responsible for heat transport. Additionally, in order to have an insightful trend of the nature of chemical bonding and electron distribution that directly affects the lattice dynamics and phonon scattering, we plotted the electron localization function. Figure \ref{phbs}(d), (e), and (f) show ELF plot of (110) plane of Cs$_2$TeCl$_6$, Cs$_2$TeBr$_6$, and Cs$_2$TeI$_6$, respectively. In these figures \ref{phbs}(d), (e) and (f), the red green and blue zone in the ELF isosurfaces around the Te-X bond exhibits the high electron localization, free electron-like feature, and low electron density region, respectively. The comparative analysis of these ELFs reveals that the ELF isosurface appears sharply localized as the red lobe is compact and tightly confined in Cs$_2$TeCl$_6$, suggesting directional covalent interaction. The localized nature of the Te-Cl bond indicates its harmonic nature. Such feature of Te-Cl makes this system have lower phonon scattering and high lattice thermal conductivity. Transitioning to Cs$_2$TeBr$_6$, we notice a subtle increase in spatial extent and diffusivities in the ELF isosurface of the Te-Br bond as the red lobe is now slightly more diffused. These diffusivities around the ELF isosurface of the Te-Br bond reflect the softening of the bond and increasing bond weakness and anharmonicity as compared to the Te-Cl bond. Such features help in greater atomic displacement and enhanced phonon-phonon interactions. This diffused nature of the red lobes in ELF isosurface of Cs$_2$TeI$_6$ becomes more prominent for the Te-I bond. The bonding region in the Te-I are more delocalized and widely spread, indicating a comparatively weak and anharmonic bond. This pronounced anharmonicity is a key contributor to phonon-phonon scattering, leading to a substantial reduction in the lattice thermal conductivity. Overall, our comparative analysis reveals that going from Cl to I as the electronegativity decreases, the electron localization decreases, and bond anharmonicity increases causing suppression of lattice thermal conductivity. These results are indicative of a better thermoelectric performance in Cs$_2$TeX$_6$ (X=Cl, Br, I). Apart from this, we have performed the ab initio molecular dynamics (AIMD) simulations in the NVT ensemble using the Nose-Hoover thermostat at 300 K, as shown in Figure S6 for Cs$_2$TeBr$_6$ and S7 for Cs$_2$TeI$_6$, the crystal framework remains intact over the long timescale of 10 ps. Throughout the simulation time, no significant structural distortion and bond breaking is observed, confirming its dynamical stability. The persistence of equilibrium over long simulation time scale indicates that these systems can maintain its integrity at 300 K.  

\subsection{Mechanical properties} 
To address the mechanical properties of these systems such as stability of materials, hardness, ductility, brittleness, etc, we have computed the elastic moduli of these materials. The elastic moduli reveal the system's response to the applied stress along different directions \cite{elastic}. As we have discussed in the section \ref{strr}, these systems have FCC structures, hence these systems have only three independent elastic moduli C$_{11}$, C$_{12}$, and C$_{44}$. The computed elastic constants are summarized in the following table \ref{tabb2}.
\begin{table}[h]
\centering
 \caption{Mechanical properties of Cs$_2$TeX$_6$ (X = Cl, Br, I) vacancy ordered perovskites. All the elastic constants and moduli are expressed in GPa.}
\begin{tabular}{ c c c c } 
 \hline
Parameters      & Cs$_2$TeCl$_6$ & Cs$_2$TeBr$_6$ & Cs$_2$TeI$_6$  \\
 \hline          
 C$_{11}$ (GPa) & 19.99      & 19.43     & 18.73 \\ 
 C$_{12}$ (GPa) & 9.93      & 9.98      & 10.49 \\ 
 C$_{44}$ (GPa) & 7.8      &  8.13      & 8.0 \\
 Shear Modulus G(GPa) & 6.54 & 6.53 & 6.12 \\
  Bulk Modulus B(GPa) & 13.29  & 13.13 & 13.23  \\
 Young Modulus Y (GPa) & 16.86 & 16.82 & 15.91 \\
  Debye temperature $\Theta_D$ (K) & 143.3 & 122 & 106.1 \\
 Longitudinal velocity $v_l$ (m/s) & 2519.9 & 2231.65 & 2111.75 \\
 Transverse velocity $v_s$ (m/s) & 1373.93 & 1220.83 & 1129.57 \\
 Average velocity $v_{av}$ (m/s) & 1532.43 & 1361.32 & 1261.64 \\
 \hline
\end{tabular}
\label{tabb2} 
\end{table}
Here, the Cauchy pressure ($C_p = C_{12} \text - C_{44}$) is positive for all systems. However, it is higher for Cs$_2$TeI$_6$ than for Cs$_2$TeCl$_6$ and Cs$_2$TeBr$_6$, indicating that Cs$_2$TeI$_6$ exhibits a greater degree of ionic character \cite{cauchy_pressure}. The shear modulus (G) quantifies a material's resistance to plastic deformation, whereas the bulk modulus (B) represents its ability to resist fracture. A B/G ratio $\geq$ 1.75 generally suggests a ductile nature, while a lower value indicates brittleness \cite{brittle}. Our calculations reveal that all these systems exhibit a predominantly ductile nature. 
The sound velocities of Cs$_2$TeX$_6$ (X=Cl, Br, I) can be estimated from the computed elastic constants \cite{sound} and the calculated values presented in the table \ref{tabb2}. The sound velocities of these systems are comparable. In particular, the Cs$_2$TeI$_6$ exhibits a lower sound velocity than Cs$_2$TeBr$_6$ and Cs$_2$TeCl$_6$. This trend in sound velocity can be attributed to the higher atomic mass of the halogen atom. Since the Debye temperature is directly proportional to the average sound velocity, the lower sound velocity of Cs$_2$TeI$_6$ results in a lower Debye temperature and becomes desirable for thermoelectric applications. \\
\subsection{Transport properties}
To assess the thermoelectric performance of these materials, we thoroughly examined their transport properties, such as electrical conductivity, thermal conductivity, Seebeck coefficient, power factor, and figure of merit. We made use of HSE06 calculated band gap along with PBE calculated electronic band structure including SOC to compute the transport properties. Our investigations predict Cs$_2$TeI$_6$ as the best thermoelectric performer compared to other perovskites Cs$_2$TeCl$_6$ and Cs$_2$TeBr$_6$ under study. At the same time, the trends for transport properties show similar behavior as function of carrier concentration and temperature for these systems. Hence, here we discuss the factors responsible for the promising thermoelectric performance of Cs$_2$TeI$_6$ in detail. The transport properties of Cs$_2$TeCl$_6$ and Cs$_2$TeBr$_6$ are given in supporting information. 
%The transport properties of Cs$_2$TeI$_6$ are discussed in detail below.

\begin{figure}[H]
 \centering
 \includegraphics[width = 0.95\textwidth]{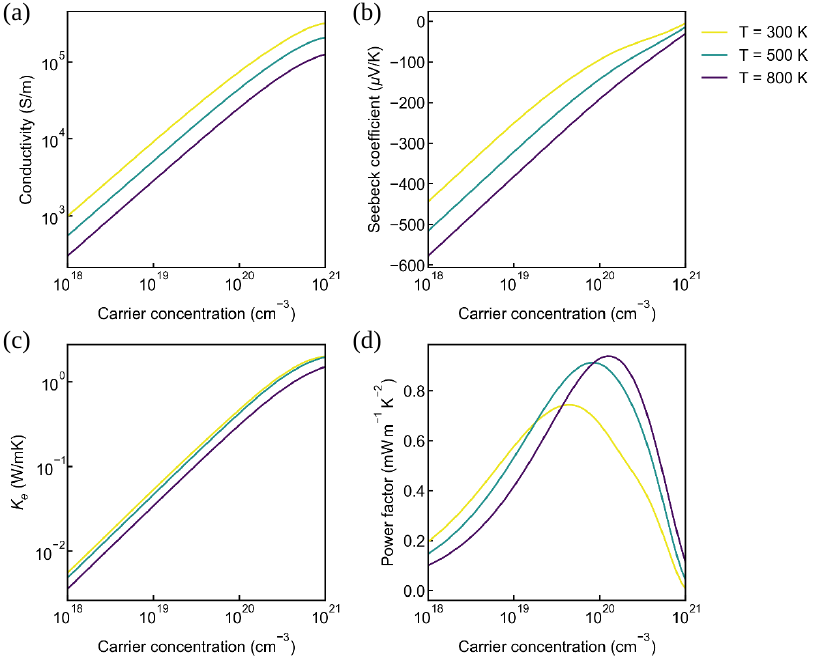}
 \caption{ Transport properties of Cs$_2$TeI$_6$ (a) electrical conductivity (b) Seebeck coefficient (c) electronic thermal conductivity (d) thermoelectric power factor for electron doping. }
\label{ntype}
\end{figure}

\begin{figure}[H]
 \centering
 \includegraphics[width = 0.95\textwidth]{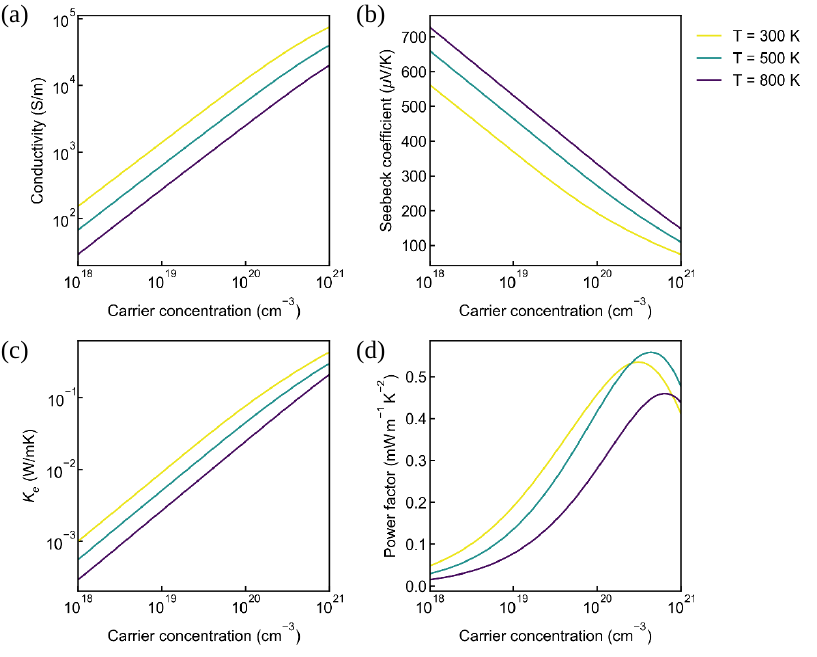}
 \caption{ Transport properties of Cs$_2$TeI$_6$ (a) electrical conductivity (b) Seebeck coefficient (c) electronic thermal conductivity (d) thermoelectric power factor for hole doping. }
\label{ptype}
\end{figure}

\begin{figure}[H]
 \centering
 \includegraphics[width = 0.9\textwidth]{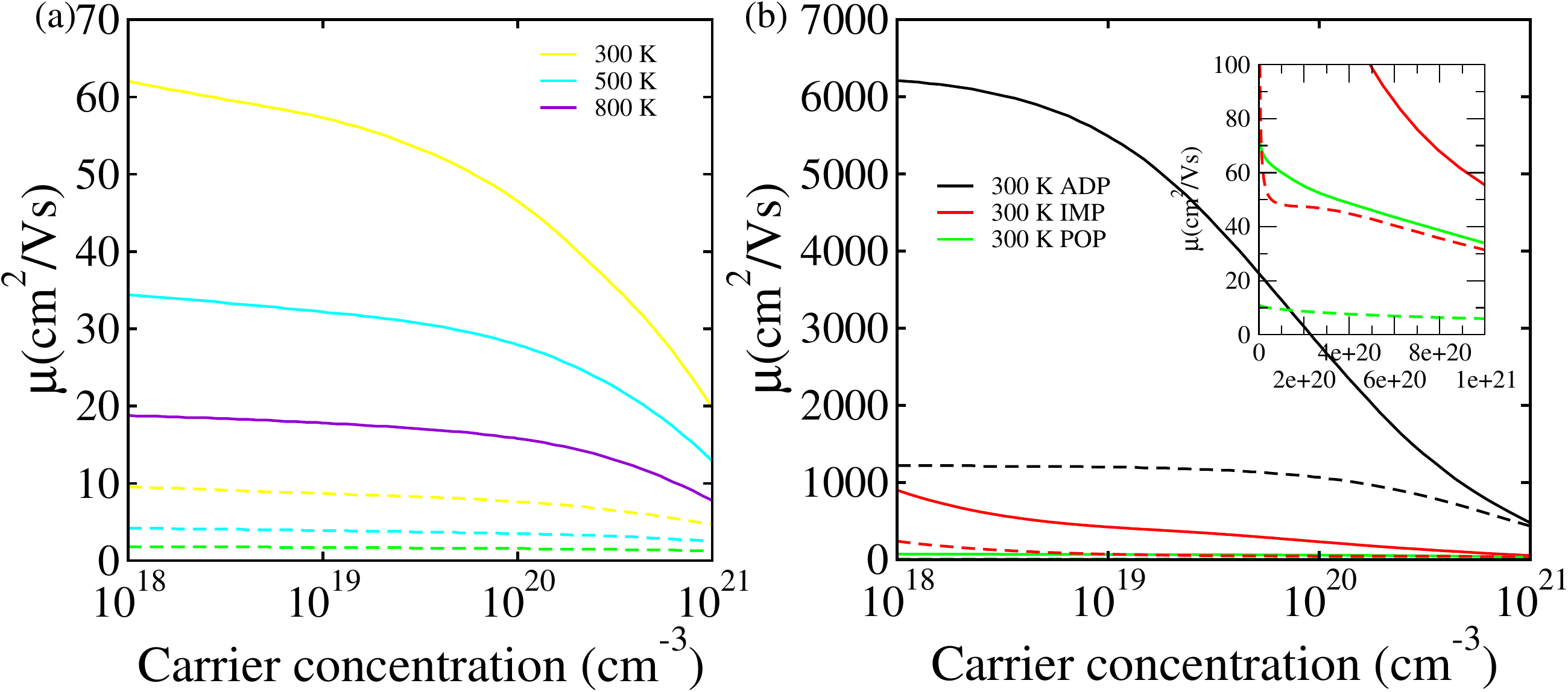}
 \caption{(a) electron and hole mobility of Cs$_2$TeI$_6$, (b) contribution of different scattering processes in overall mobility of Cs$_2$TeI$_6$ for a range of temperature from 300 K to 800 K. The solid and dotted lines indicate electron and hole doping, respectively.}
\label{mob}
\end{figure}

 \noindent \paragraph*{Electrical conductivity} 
The electrical conductivity ($\sigma$) of Cs$_2$TeI$_6$ for electron and hole doping as a function of carrier concentration over a temperature range of 300 K to 800 K is presented in Figures \ref{ntype}(a) and \ref{ptype}(a), respectively. The electrical conductivity is significantly higher for electron doping than for hole doping. This behavior is attributed to the presence of light carriers in highly dispersive conduction bands, in contrast to the heavy carriers present in flat valence bands, as shown in the table \ref{stdd}.
At a low electron (hole) doping concentration of $1 \times 10^{18}$ cm$^{-3}$ and at 300 K, the electrical conductivity is found to be $9.93 \times 10^{2}$ $\textrm Sm^{-1}$ ($1.53 \times 10^{2}$ $\textrm Sm^{-1}$). With increasing doping concentration, the electrical conductivity monotonically increases due to the semiconducting nature of these materials but decreases at high carrier concentrations due to an increased scattering rate. At a high electron (hole) doping concentration of $1 \times 10^{21}$ cm$^{-3}$ and at 300 K, $\sigma$ is found to be $3.17 \times 10^{5}$ $\textrm Sm^{-1}$ ($7.50 \times 10^{4}$ $\textrm Sm^{-1}$) for Cs$_2$TeI$_6$. With increase in temperature, $\sigma$ reduces to be $3 \times 10^{2}$  $\textrm Sm^{-1}$ ($2.89 \times 10^{1}$  $\textrm Sm^{-1}$) at electron (hole) concentration of $1 \times 10^{18}$ cm$^{-3}$ and at 800 K. This decrease in $\sigma$ is caused by enhanced scattering at high temperatures.
This decline in electrical conductivity is mainly attributed to decrease in carrier mobility at high temperatures and high carrier concentration. Figure \ref{mob}(a) shows the carrier mobility as a function of carrier concentration for a range of temperature for Cs$_2$TeI$_6$. In the figure \ref{mob}, the solid and dashed lines represent electron and hole doping, respectively. To delve deeper into the mobility limiting factors, we looked into the contributions of acoustic deformation potential (ADP), polar optical phonon (POP) , and impurity (IMP) caused scattering, as depicted in figure \ref{mob} (b) at room temperature for electron and hole mobility. Our analysis reveals that IMP and POP mechanisms are the most dominant scattering processes in limiting mobility in comparison with the ADP scattering process. As shown in figure \ref{mob}(b), when IMP and POP scattering contributions are smaller, total mobility is mainly limited by these mechanisms, as described by Matthiessen's rule \cite{matthiessen}. Near room temperature, POP scattering is the dominant contributor, whereas IMP scattering becomes more significant at higher temperatures for both electrons and holes. Holes in Cs$_2$TeI$_6$ exhibit significantly lower mobility than electrons, resulting in lower electrical conductivity for holes. Remarkably, the electrical conductivity of Cs$_2$TeI$_6$ is comparable to state of the art thermoelectric materials, namely SnSe (1000 $\textrm Sm^{-1}$ at 323 K) \cite{snse}, SnS (500 $\textrm Sm^{-1}$ at 323 K) \cite{sns}, and Cu$_2$Se (1500 $\textrm Sm^{-1}$ at 323 K) \cite{cu2se}. 

\noindent \paragraph*{Seebeck coefficient}
To evaluate the potential difference produced by the thermal gradient, we evaluated the Seebeck coefficient (S) of Cs$_2$TeI$_6$ as a function of carrier concentration for electron and hole doping, over a temperature range of 300 K to 800 K as shown in figures \ref{ntype}(b) and \ref{ptype}(b), respectively. As opposed to electrical conductivity, the Seebeck coefficient is consistently greater for hole doping as compared to electron doping across the considered doping concentration range owing to high effective mass of holes as shown in the table \ref{stdd}. For instance, at an electron (hole) concentration of $1 \times 10^{18}$ cm$^{-3}$ and 300 K, the Seebeck coefficient is found to be 440 (562) $\mu \textrm VK^{-1}$ for Cs$_2$TeI$_6$. This difference is caused by the significantly heavier holes as compared to the electrons. This difference in magnitude of the Seebeck coefficient becomes more prominent with increase in temperature and 800 K, the Seebeck coefficient increases to be 577 (728) $\mu \textrm VK^{-1}$. The Seebeck coefficient values obtained for these materials are notably comparable to those reported for other double perovskites, such as Rb$_2$AgBiX$_6$​ (X = Cl, Br) \cite{rb2agbix6}, K$_2$​AgAsX$_6$​ (X = halogen elements) \cite{k2agasx6}, and Cs$_2$​BiAgX$_6$​ (X = halogen elements) \cite{cs2biagx6}, K$_2$InBiX$_6$ (X=Cl, Br) \cite{k2inbix6}, CH$_3$NH$_3$PbI$_3$ \cite{ch3nh3pbi3} as well as the commercially available Bi$_2$​Te$_3$​ \cite{bi2te3-cb}. Increasing the doping concentration lowers the Seebeck coefficient further and at an electron (hole) concentration of $1 \times 10^{18}$ cm$^{-3}$ and 300 K, it becomes 5 (74) $\mu \textrm VK^{-1}$ and improves slightly with increase in temperature. These effects are in compliance with the Pisarenko relation \cite{pisarenko}. These high values of the Seebeck coefficient for Cs$_2$TeI$_6$ show this system's strong potential for thermoelectric applications.

\noindent \paragraph*{Electronic thermal conductivity}
Figure \ref{ntype}(c) and \ref{ptype}(c) display the electronic thermal conductivity of Cs$_2$TeI$_6$ as a function of carrier concentration over a temperature range of 300 K to 800 K for electron and hole doping, respectively. As $\sigma$ and $\kappa_e$ are proportional to each other by Wiedemann-Franz law, $\kappa_e$ follows the similar trend like $\sigma$ in this system Cs$_2$TeI$_6$ for both electron and hole doping. Like $\sigma$, the $\kappa_e$ increases with increase in carrier concentration and decreases with temperature as mobility decreases with increase in temperature (Figure \ref{mob}). For instance, at an electron (hole) doping concentration of $1 \times 10^{18}$ cm$^{-3}$ and 300 K, $\kappa_e$ is calculated to be $5.57 \times 10^{-3}$ ($9.97 \times 10^{-4}$) $\textrm Wm^{-1}K^{-1}$. These values are ultralow and favorable to design efficient thermoelectric devices. With increase in doping concentration, at $1 \times 10^{21}$ cm$^{-3}$, $\kappa_e$ enhances to be 2.02 (0.4) $\textrm Wm^{-1}K^{-1}$. Further with increase in temperature, it diminished to be 1.51 (0.2) $\textrm Wm^{-1}K^{-1}$. Such low values motivates us further to investigate the thermoelectric performance of this system Cs$_2$TeI$_6$.

\noindent \paragraph*{Thermoelectric power factor} 
With the above calculated transport properties, we evaluate the power factor ($S^2\sigma$) in order to quantify its capability in power generation. Figure \ref{ntype}(d) and \ref{ptype}(d) show the thermoelectric power factor of Cs$_2$TeI$_6$ as a function of carrier concentration for temperatures ranging from 300 K to 800 K for electron and hole doping, respectively. We reveal that the power factor initially increases with increase in carrier concentration, achieves maxima and then decreases with further increase in carrier concentration. This behavior stems from the trade-off behavior between the Seebeck coefficient and electrical conductivity. We find that the power factor is significantly higher for electron doping as compared to hole doping owing to increased electrical conductivity for electron doping. Our calculations reveal that the power factor maximizes to be 0.743 (0.535) $\textrm mWm^{-1}K^{-2}$ at optimum electron (hole) carrier concentration of $4.23 \times 10^{19}$ ($3.10 \times 10^{20}$) cm$^{-3}$ and at 300 K. These values improve further with increase in temperature. For instance, the maximum power factor for electron doping is obtained to be 0.94 at electron concentration of $1.21 \times 10^{20}$ cm$^{-3}$ at 800 K, whereas for hole doping, the maximum calculated power factor is 0.56 $\textrm mWm^{-1}K^{-2}$ at hole concentration of $6.26 \times 10^{20}$ cm${-3}$ and at 500 K. These values are comparable to some state of the art thermoelectric materials, such as 
PbTe(2.5 $\textrm mWm^{-1}K^{-2}$) \cite{pfpbte_data}, Bi$_2$Te$_3$ (n-type: $4.5$ $\textrm mWm^{-1}K^{-2}$, p-type: $3.0$ $\textrm mWm^{-1}K^{-2}$)\cite{pfbi2}, SnSe(0.8 $\textrm mWm^{-1}K^{-2}$)\cite{pfsnse_data}, Bi$_2$Se$_3$(2.7 $\textrm mWm^{-1}K^{-2}$) \cite{pfbi2se3_data}, Yb$_{14}$MnSb$_{11}$ (0.6 $\text mWm^{-1}K^{-2}$) \cite{pfyb}, CoSb$_3$(1.6 $\textrm mWm^{-1}K^{-2}$) \cite{pfcosb3_data}, CsCdBr$_6$ (0.98 $\textrm mWm^{-1}K^{-2}$) \cite{pfcscdbr6_data}, CsCdCl$_6$ (0.41 $\textrm mWm^{-1}K^{-2}$) \cite{pfcscdbr6_data}, and K$_2$OsBr$_6$ (0.06 $\textrm mWm^{-1}K^{-2}$) \cite{pfk2osbr6_data}.

\begin{figure}[H]
 \centering
  \includegraphics[width=0.9\textwidth]{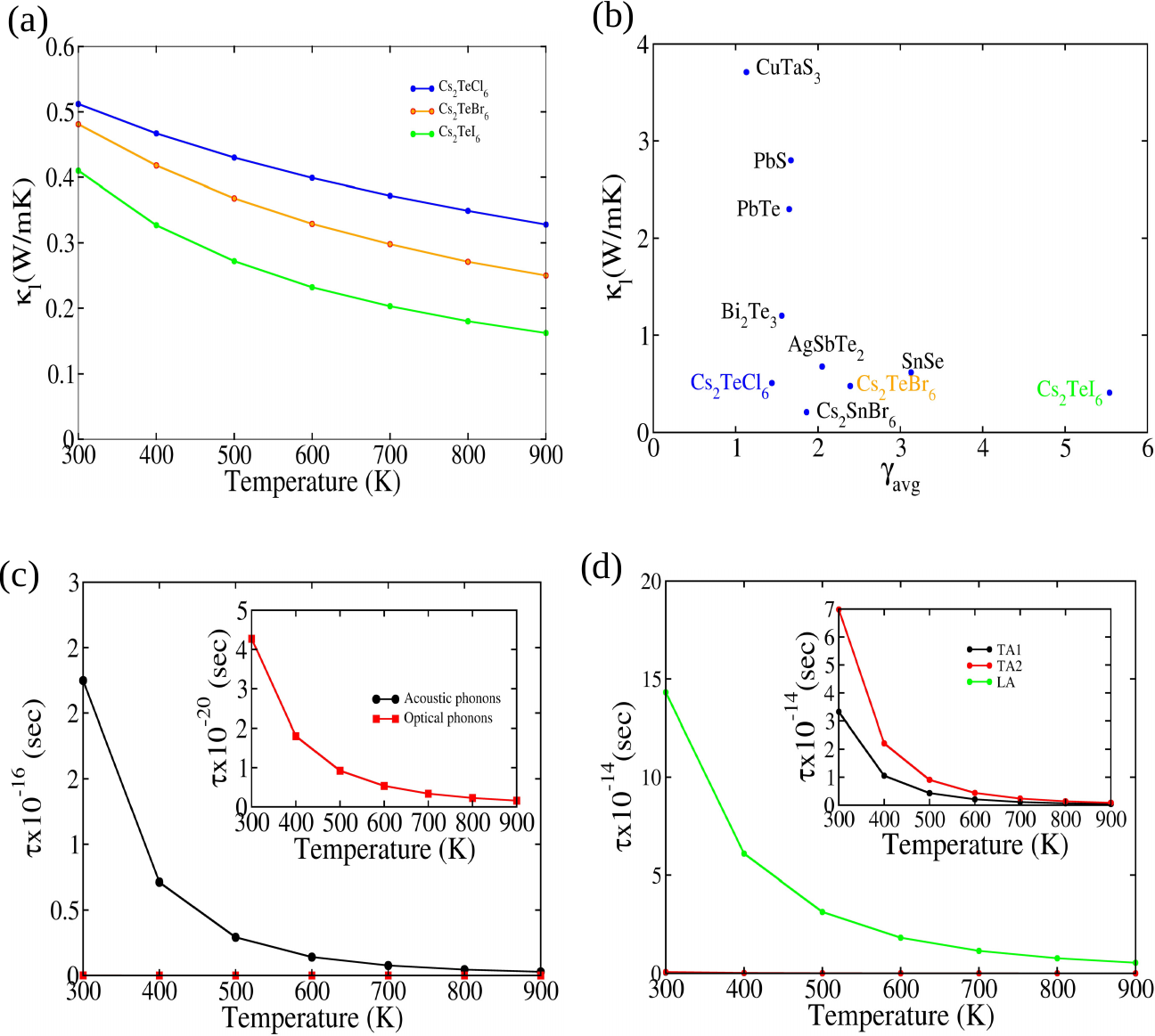}
 \caption{(a) Lattice thermal conductivity ($\kappa_l$) as a function of the temperature of Cs$_2$TeX$_6$(X=Cl, Br, I), (b) Average Gr\"uneisen parameter of known thermoelectric materials, (c) Relaxation time of acoustic and optical phonons, (d) Mode specific relaxation time of acoustic branch of Cs$_2$TeI$_6$}
\label{ltc}
\end{figure}

\begin{figure}[H]
 \centering
  \includegraphics[width=0.6\textwidth]{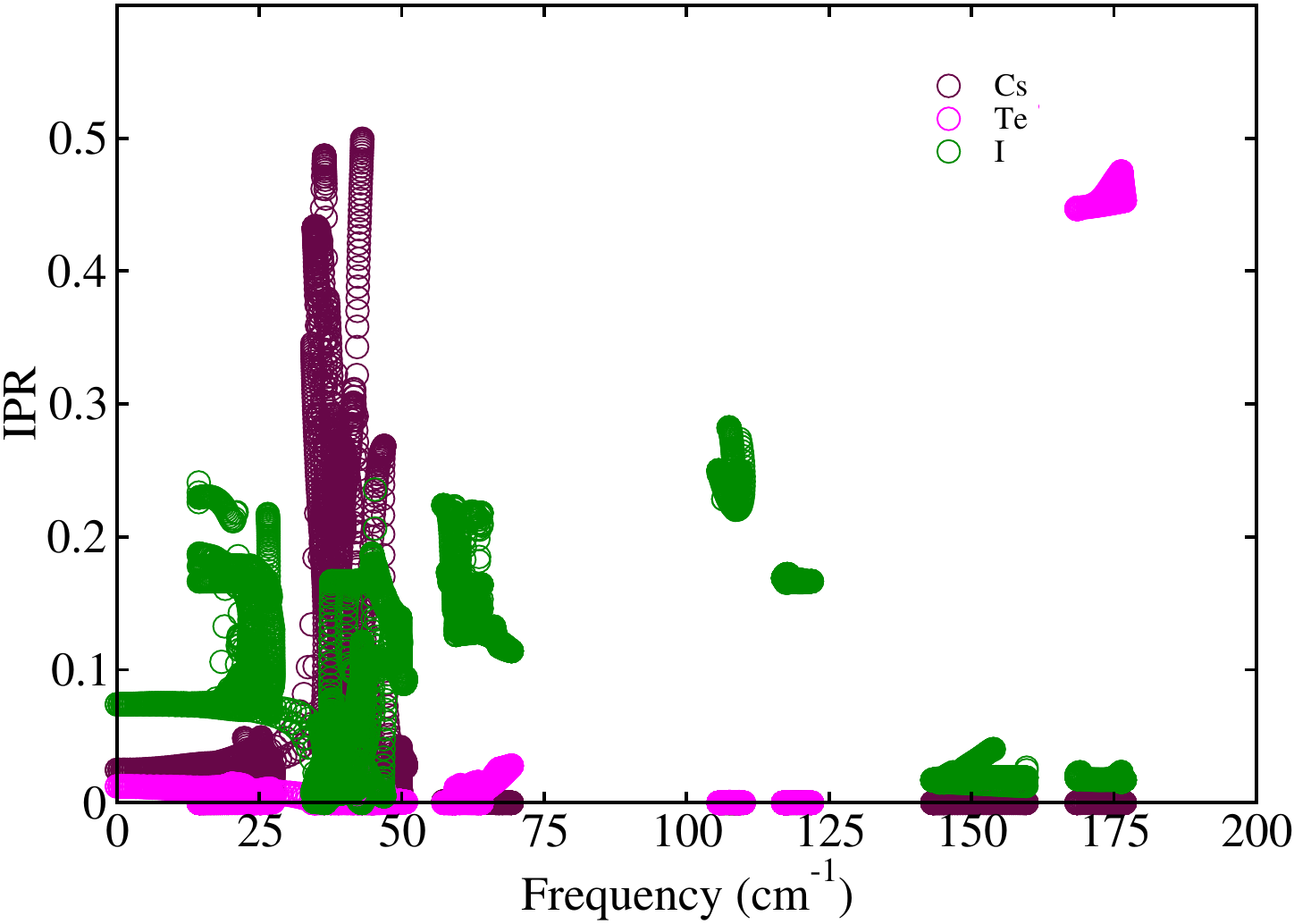}
 \caption{Inverse participation ratio as a function of frequency of Cs$_2$TeI$_6$}
\label{iprimg}
\end{figure}

\begin{figure}[H]
 \centering
  \includegraphics[width=0.8\textwidth]{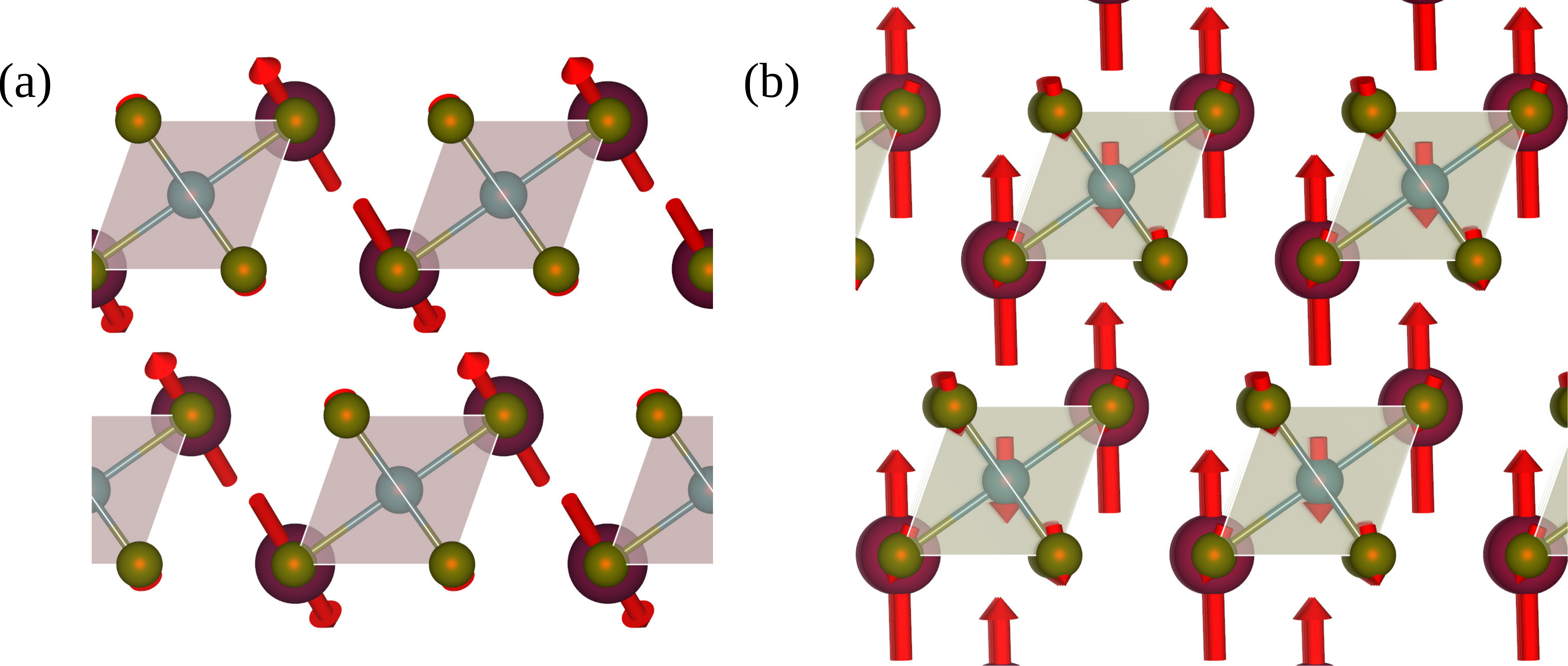}
 \caption{Displacement pattern associated to (a) 34 cm$^{-1}$  and (b) 42 cm$^{-1}$ for Cs$_2$TeI$_6$}
\label{iprmodes}
\end{figure}

\paragraph*{Lattice thermal conductivity}
Figure \ref{ltc}(a) displays the lattice thermal conductivity ($\kappa_l$) as a function of temperature for Cs$_2$TeI$_6$. $\kappa_l$ decreases as temperature increases following the temperature dependence as T$^{-1}$ \cite{umklapp} as shown in the figure S1. This type of temperature variance with $\kappa_l$ implies the dominance of Umklapp scattering in this system. $\kappa_l$ was found to be 0.41 $\textrm Wm^{-1}K^{-1}$ at room temperature. Such low values at room temperature are beneficial for better thermoelectric performance. The calculated $\kappa_l$ is much lower than some well known thermoelectric materials namely, Bi$_2$Te$_3$(1.24 $\textrm Wm^{-1}K^{-1}$) \cite{klbi}, SnSe (1.19 $\textrm Wm^{-1}K^{-1}$) \cite{klsnse}, Y$_2$Te$_3$ (2.96 $\textrm Wm^{-1}K^{-1}$) \cite{klyt}, and PbTe (2.00 $\textrm Wm^{-1}K^{-1}$) \cite{klpb}. Such perovskite materials are well known for low lattice thermal conductivity owing to their complex structure and weak bonds. In general, the lattice thermal conductivity of a solid depends upon three parameters namely, specific heat (C$_v$), group velocity (v$_g$), and relaxation time ($\tau$) as given in the equation \ref{lateq}. The v$_g$ and $\tau$ parameters are the main deciding factors behind $\kappa_l$ since C$_v$ remains constant at room temperature and above. The macroscopic group velocity in Cs$_2$TeI$_6$ is significantly low as given in table \ref{tabb2}. Additionally, the low bulk modulus of these system suggest weaker bonds as given in the table \ref{tabb2} and weak bonds are more anharmonic and cause low lattice thermal conductivity \cite{anharmonic}. This is also in compliance with our projected crystal orbital Hamilton population as discussed in section: \ref{cohps}. In addition to this, figure \ref{ltc} (b) compares the average Gr\"uneisen parameter of Cs$_2$TeI$_6$ with other known thermoelectric materials. A larger Gr\"uneisen parameter of Cs$_2$TeI$_6$ indicates the lower lattice thermal conductivity \cite{anharmonic}. Importantly, this value subsides further with increase in temperature and reaches to be 0.18 $\textrm Wm^{-1}K^{-1}$ at 800 K. Scattering rate, especially phonon-phonon scattering becomes the dominating factor in lattice thermal conductivity at high temperatures as phonon population significantly enhances at high temperatures. There are three types of phonon-phonon scattering namely, normal scattering ($\tau_N$), Umklapp scattering ($\tau_{umk}$) and isotropic scattering ($\tau_{iso}$). These scattering processes are caused by acoustic and optical phonons. In order to have a clear picture of these scattering mechanism, we plotted these relaxation time as function of temperature as given in the figure \ref{ltc}(c). Figure \ref{ltc}(c) displays the total relaxation time of acoustic and optical phonons. The total relaxation time is calculated using equation \ref{taueq} as given in the Mattheissen's rule \cite{matthiessen}. Though optical phonons scattering rate is significantly lower than acoustic phonons scattering, it does not directly affect the $\kappa_l$ but offers resistance to acoustic phonons. This is also evident from the significant overlap of acoustic and optical modes in the figure \ref{phbs}. Notably, the relaxation time of acoustic phonons is lower too as compared to relaxation time in isotropic semiconductors. Further, we displayed the relaxation time of different acoustic modes in the figure \ref{ltc}(d). Figure \ref{ltc}(d) demonstrates that the relaxation time associated with transverse acoustic (TA) mode is significantly lower than the longitudinal acoustic (LA) modes. In particular, the first transverse acoustic modes (TA1) significantly limits the $\kappa_l$ of Cs$_2$TeI$_6$. We further analysed the inverse participation ratio (IPR) to figure out the real cause for such ultralow $\kappa_l$ at the microscopic level. The IPR is defined in the equation \ref{ipreq}, where N denotes the total number of atoms and $u_{i\alpha q}$ indicates normalized eigenvector for $i^{th}$ atom along $\alpha$ direction and for $q^{th}$ phonon mode. The IPR helps us to recognize the particular  mode that contributes to ultralow $\kappa_l$. Figure \ref{iprimg} depicts the projected IPR vs frequency for the constituent atoms in the system Cs$_2$TeI$_6$. Our calculation reveals that Cs has highest contribution to IPR suggesting that the Cs atom is actively participating in scattering the phonons. Here we are neglecting the crucial contribution of Te to IPR at high frequency as these high frequency phonons do not play significant role in transport. The Cs dominated highest IPR is associated to particular modes around low frequencies 34 cm$^{-1}$ and 42 cm$^{-1}$. The displacement pattern of these modes at $\sim$ 34 cm$^{-1}$ and 42 cm$^{-1}$ are displayed in figures \ref{iprmodes}(a) and (b), respectively. Figure \ref{iprmodes}(a) displays that the TeI$_6$ octahedra shows tilting and asymmetric distortions along with neighbouring Cs atoms vibrating in  opposite directions with different amplitudes. Such non-uniform and out-of-phase displacement modes break the periodicity required for phonon conduction, resulting into strong phonon scattering, thereby reduces the phonon mean free path and makes phonon less effective in carrying heat. Furthermore, figure \ref{iprmodes} (b) represents the  Cs atoms exhibiting large amplitude vertical displacement, while TeI$_6$ octahedral network remains relatively rigid. Such displacement pattern confirms that Cs acts as a rattler ion inside a soft cage formed by TeI$_6$ octahedral network. These low frequency Cs dominated vibrations strongly hinder the heat carrying acoustic phonons. Overall, the collective tilting of Te-I units and rattling like Cs vibrations turn the lattice network into phonon scattering network that strongly limits the phonon conduction. Here it is worth mentioning that these ultralow $\kappa_l$ values are much desirable for efficient thermoelectric performance.

\begin{equation}
\kappa_l=\frac{1}{3} C_v v_g^2 \tau 
\label{lateq}
\end{equation}

\begin{equation}
\tau^{-1}= \tau_N^{-1} + \tau_{umk}^{-1} + \tau_{iso}^{-1}
\label{taueq}
\end{equation}

\begin{equation}
\mathrm{IPR}= \sum_{i=1}^{N} \left( \sum_{\alpha=1}^{3} u_{i\alpha,q}^{2} \right)^{2}
\label{ipreq}
\end{equation}

\begin{figure}[H]
 \centering
  \includegraphics[width=0.50\textwidth]{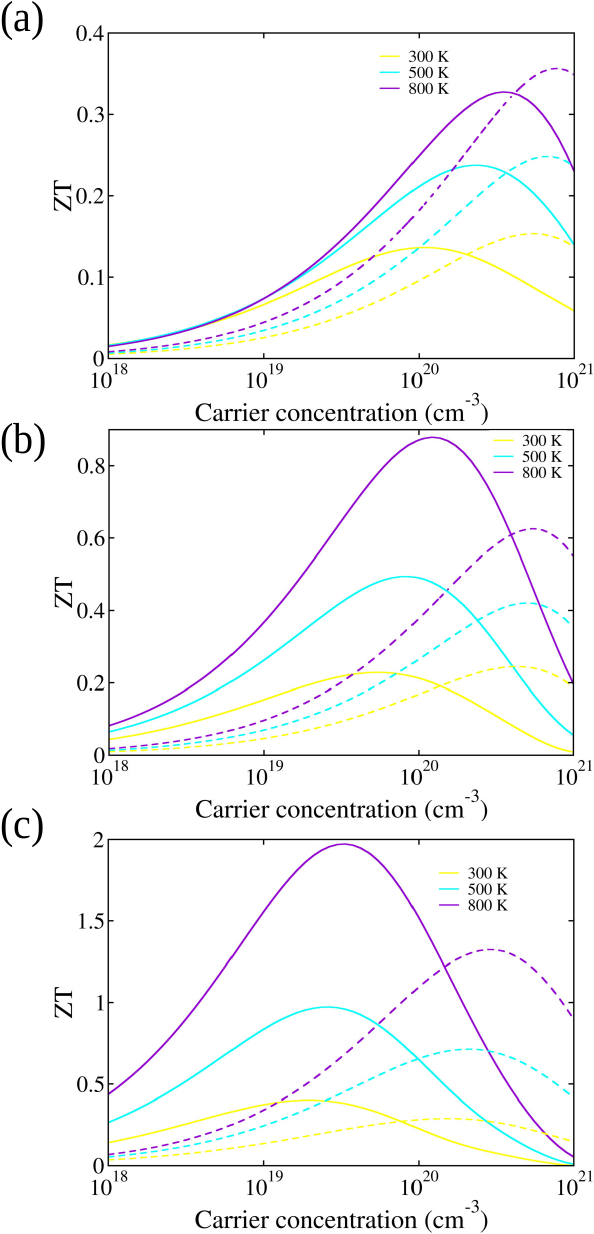}
 \caption{Figure of merit of (a) Cs$_2$TeCl$_6$ (b) Cs$_2$TeBr$_6$ (c) Cs$_2$TeI$_6$ as a function of carrier concentration for 300 K, 500 K, and 800 K}
\label{ztt}
\end{figure}

\paragraph*{Figure of Merit} 
Figure \ref{ztt} illustrates the figure of merit as a function of carrier concentration and for a range of temperatures. Here, the solid lines and dashed lines indicate electron and hole doping, respectively. The highest ZT among Cs$_2$TeX$_6$(X=Cl, Br, I) was found to be 1.97 at optimized carrier concentration $3.35 \times 10^{19}$ cm$^{-3}$ and 800 K for n type doping in Cs$_2$TeI$_6$. However, for p-type doping in Cs$_2$TeI$_6$, the highest ZT was achieved to be 1.31 at hole concentration $2.75 \times 10^{20}$ cm$^{-3}$ and 800 K. High Seebeck coefficient and electrical conductivity along with low thermal conductivity contributes to this elevated ZT in Cs$_2$TeI$_6$ at such a low carrier concentration. Essentially, the suitable band curvature in the conduction band of Cs$_2$TeI$_6$ along with the presence of octahedra in the local crystal structure are the reason behind such a high ZT. This ZT is higher than that of some state of the art thermoelectric materials namely, Bi$_2$Te$_3$ (~1.03 at 400 K)\cite{ztbi2te3}, Bi$_2$Se$_3$ (1.14 at 300 K) \cite{pfbi2se3_data}, CoSb$_3$(0.43 at 600 K) \cite{pfcosb3_data}, SnTe (1 at 900 K) \cite{klsnte1}, PbTe (1.96 at 700 K) \cite{ztpbte}, SiGe (1.84 at 1100 K) \cite{pfsige}, CsCdBr$_6$ (1.16 at 900 K) \cite{pfcscdbr6_data} and Cs$_2$InAgCl$_6$ (0.74 at 700 K) \cite{cs2inagcl6}. Among the perovskite family, the ZT is significantly higher than that of many prospective modern latest thermoelectric materials, including, CeMnO$_3$(0.40) \cite{remno3}, PrMnO$_3$ (0.47 at 500 K) \cite{remno3}, CH$_3$NH$_3$PbI$_3$ (0.7) \cite{ztch3nh3pbi3}, BiGaO$_3$ (0.62) \cite{bigao3}, SrTiO$_3$ (0.18) \cite{srtio3}, and Cs$_2$PdCl/Br$_6$ (0.71/0.76) \cite{cs2pdx6}.

\section{Conclusions} 
\label{conc}
To summarize, we carried out systematic calculations to predict the thermoelectric performance of Cs$_2$TeX$_6$ (X=Cl, Br, I) with both electron and hole doping. Our calculations reveal that these systems Cs$_2$TeX$_6$ (X=Cl, Br, I) have cubic structure with electronic band gap of 3.27, 2.50, and 1.55 eV, respectively. These systems are mechanically and dynamically stable. We explored the effect of doping concentration upon the transport properties namely, Seebeck coefficient, electrical and thermal conductivity using the rigid band approach. We predict that Cs$_2$TeI$_6$ has better thermoelectric performance than other systems in our study. With electron doping of $3.35 \times 10^{19}$ cm$^{-3}$, the figure of merit was found to be 1.97 at 800 K, whereas highest ZT is 1.31 for hole doping at hole concentration $2.75 \times 10^{20}$ cm$^{-3}$ and at 800 K. This high ZT of this system stems from the suitable conduction band dispersion along with weak bonds in local structure that gives low macroscopic phonon velocity, in turn, ultralow lattice thermal conductivity. Such a high ZT at a low carrier concentration makes these materials unique in this area. Overall, the synergistic interplay between favorable band dispersion and ultralow lattice thermal conductivity renders Cs$_2$TeI$_6$ as a strong candidate for high performance thermoelectric applications.  
 
\section{Acknowledgements} 
Authors gratefully acknowledge the computational resources provided by the Paramrudra High-Performance Computing (HPC) facility at the Inter-University Accelerator Centre (IUAC) New Delhi, C-DAC pune, IIT Jodhpur, and Sharda university Agra. The authors used ChatGPT to review the sentence structure and edited the content as required. 
\section{Declaration of competing interest}
The authors claim no competing interest to report.
\bibliographystyle{unsrt}
\bibliography{cstex_ref}

\end{document}